\documentclass[pre,twocolumn,showpacs,amssymb,superscriptaddress]{revtex4}
\usepackage{graphicx}

\begin{document}

\title{Asymptotic Scaling of the  Diffusion Coefficient of Fluctuating
``Pulled'' Fronts}

\author{Debabrata Panja} \affiliation{Institute for Theoretical
Physics, Universiteit van Amsterdam, Valckenierstraat 65, 1018 XE
Amsterdam, The Netherlands}

\begin{abstract} 
We present a (heuristic) theoretical derivation for the scaling of the
diffusion coefficient $D_f$ for fluctuating ``pulled'' fronts. In
agreement with earlier numerical simulations, we find that as
$N\rightarrow\infty$, $D_f$ approaches zero as $1/\ln^3N$, where $N$
is the average number of particles per correlation volume in the
stable phase of the front. This behaviour of $D_f$ stems from the
shape fluctuations at the very tip of the front, and is independent of
the microscopic model.
\end{abstract} 


\maketitle

Interest in the effect of fluctuations on propagating fronts has
revived in recent years with the understanding that fronts which in
the deterministic mean-field limit are of the  so-called {\em pulled}
type are surprisingly sensitive to fluctuation and discrete-particle
effects. Pulled fronts are those that propagate  into a linearly
unstable state, and whose asymptotic front speed $v_{\text{as}}$ is
simply the linear spreading speed $v^*$ of infinitesimal perturbations
around the linearly unstable state \cite{dee1,dee2,ebert}. The
propagation mechanism of such fronts is that they are being ``pulled
along'' by the growth and spreading of small perturbations around the
linearly unstable state. It was first established by Brunet and
Derrida \cite{bd} and later confirmed in a variety of stochastic front
equations \cite{bd,vanzon,kns,levine,bd3} that  when pulled fronts are
realized by stochastic moves of discrete particles on a lattice, such
that the average number of particles per lattice site or correlation
volume in the saturation phase of the front is $N$, $v_{\text{as}}$
approaches $v^*$ from below extremely slowly: the convergence to $v^*$
scales only as $1/\ln^2N$ when $N\rightarrow\infty$, with a known
prefactor that depends on the model under consideration. The reason
why a true pulled front is so sensitive to finite particle cutoff
effects is the very fact that there is essentially no growth below the
cutoff  level of one ``quantum'' of particle. As a result, these
discrete particle front realizations  are actually weakly pushed
\cite{PvS1,PvS2,PvS3}. To remind ourselves that they converge to
pulled  fronts in the limit $N\to\infty$, here we will refer  to them
as fluctuating ``pulled'' fronts.

The second important feature of fluctuating ``pulled'' fronts, namely
the diffusive wandering of the front itself around its average
position as a result of stochasticity in the microscopic dynamics,
still remains very poorly understood. Of particular interest is the
question how the front diffusion coefficient $D_f$ vanishes as
$N\rightarrow\infty$. This scaling is particularly difficult to study
numerically. Nevertheless, using a clever algorithm to take $N$ as
large as $10^{150}$, Brunet and Derrida \cite{bd3} presented
convincing numerical evidence that for the model they studied, $D_f$
scales as $\sim1/\ln^3N$. Moreover, a ``simplified model'', where the
fluctuations are {\it randomly\/} generated {\it only\/} on the
instantaneous foremost occupied lattice site (i.f.o.l.s.), was found
to exhibit the {\em same} $1/ \ln^3 N$ asymptotic scaling of $D_f$ as
the full stochastic model \cite{bd3}.

The microscopic dynamics of Brunet and Derrida's (discrete-time) model
\cite{bd3} closely resembles that of the  (continuous-time) ``clock
model'' \cite{vanzon}. In both models, one considers a set of $N$
particles with integral ``readings'' $k=0,1,2,\ldots$. The number of
particles with a certain reading $k$ at time $t$ is $n_k(t)$. With
$\phi_k(t)=\sum_{k'=k}^{\infty}n_{k'}(t)/N$, in a computer simulation,
the speed and the diffusion coefficient of the front in both models
are measured by tracking the position of the centre of mass of the
particle distribution (henceforth referred to as the centre of mass of
the front itself) in individual realizations. For the particle
distribution $\{n_k(t)\}_{\text r}$ of realization r at time $t$, in
both models, the centre of mass of the front is located at
\begin{eqnarray}
S_{\text r}(t)\,=\,N^{-1} \sum_k k n_k(t)=\sum_k\phi_k(t)
\label{srt}
\end{eqnarray}
From there onwards, one defines
$D_f=\lim_{T\rightarrow\infty}d\langle[S_{\text r}(t+T)-S_{\text
r}(t)-v_NT]^2\rangle/dT$ and $v_N=\langle\dot{S}_{\text
r}(t)\rangle$. The angular brackets denote first an average over all
possible updating random number sequences for a given front
realization {\it after\/} time $t\gg1$, and then a further average
over the ensemble of (initial) front realizations {\it at\/} time $t$.

To the best of our knowledge, at present, there exists no analytical
derivation of the $1/\ln^3N$ asymptotic scaling of $D_f$ for
fluctuating ``pulled'' fronts. Our aim here is to provide a derivation
for the stochastic Fisher-Kolmogorov-Petrovsky-Piscunov (sFKPP)
equation
\begin{equation}
\partial_t\phi\,=\,\partial_x^2\phi\,+\,\phi\,-\phi^2\,+\,\sqrt{\phi\,-\phi^2}\,\eta(x,t)/\sqrt{N}\,,
\label{fkpppart}
\end{equation}
where the stochastic term $\propto\eta(x,t)$ is interpreted in the
It\^{o} sense with the two following conditions:
\begin{eqnarray}
\langle \eta(x,t)\rangle_\eta=0,\,\langle
\eta(x,t)\eta(x',t')\rangle_\eta\!=\delta(x-x')\,\delta(t-t').
\label{ito2}
\end{eqnarray}
We then argue that the same scaling of $D_f$ holds for the clock model
\cite{vanzon} and for the microscopic model that Brunet and Derrida
considered for their simulation \cite{bd3} --- in other words, the
asymptotic scaling of $D_f$ is independent of the microscopic model
and is a generic property of fluctuating ``pulled'' fronts. The full
flavour of this subtlety-riddled derivation appears elsewhere
(Secs. 2.5 and 4.2, Ref. \cite{review}). Here we focus only on the
main points and the main results.

From this perspective, it is therefore important to first summarize
the Langevin-type field-theoretical approaches for general
reaction-diffusion systems \cite{review,lutz1,rocco1,armero}. Starting
with the It\^{o} stochastic differential equation
\begin{equation}
\partial_t\phi\,=\,\partial_x^2\phi\,+\,f(\phi)\,+\,\tilde\varepsilon^{1/2}R(x,t)\,,
\label{fkpp}
\end{equation}
with $\tilde\varepsilon\ll1$ and $R(x,t)=g(\phi)\eta(x,t)$, in these
approaches, one writes the corresponding front solution as
$\phi(x,t)=\phi^{(0)}[x-v_{\text as}t-X(t)]+\delta\phi[x-v_{\text
{as}}t-X(t)]$ to separate the systematic and the fluctuating part of
the front. The systematic part $\phi^{(0)}(\xi)$ satisfies the
deterministic equation $-v_{\text
{as}}\partial_\xi\phi^{(0)}(\xi)=\partial_\xi^2\phi^{(0)}(\xi)+f[\phi^{(0)}(\xi)]$.
As for the fluctuating part, the idea behind writing $\phi(x,t)$ in
the above form is to separate the fluctuations in the front at two
different time scales. Of these, the long time scale fluctuations are
coded in the random wandering $X(t)$ of the Goldstone mode
$\Phi_{G,R}(\xi)\equiv d\phi^{(0)}(\xi)/d\xi$ of the front around its
uniformly translating position $x-v_{\text{as}}t$. On the other hand,
the short time scale fluctuations manifest themselves through the
fluctuations in the front shape around its instantaneous position
$x-v_{\text{as}}t-X(t)$.

In this form, the Goldstone mode is in fact the right eigenvector of
the linear stability operator ${\cal
L}_{v_{\text{as}}}=\partial_\xi^2+v_{\text{as}}\partial_\xi+\delta_\phi
f(\phi)|_{\phi=\phi^{(0)}}$ with zero eigenvalue \cite{PvS1}; and its
instantaneous position is defined by requiring that it be orthogonal
to the front shape fluctuations, i.e.,
$\int_{-\infty}^{\infty}d\xi\,\Phi_{G,L}(\xi)\delta\phi[\xi-X(t),t]=0$
at all times \cite{review,lutz1,rocco1,armero}. Here,
$\Phi_{G,L}(\xi)\equiv e^{v_{\text{as}}\xi}\Phi_{G,R}(\xi)$ is the
left eigenvector of ${\cal L}_{v_{\text{as}}}$ corresponding to
eigenvalue zero \cite{strict}. When this orthogonality condition is
used in linearized Eq. (\ref{fkpp}), one finds that the  instantaneous
speed of the Goldstone mode fluctuates (at long time  scales) around
$v_{\text{as}}$ by \cite{review,lutz1,rocco1,armero}
\begin{eqnarray}
\dot{X}(t)\,=\,-\,\tilde\varepsilon^{1/2}\frac{\int_{-\infty}^{\infty}d\xi\,\Phi_{G,L}(\xi)\,R(\xi,t)}{\int_{-\infty}^{\infty}d\xi\,\Phi_{G,L}(\xi)\,\Phi_{G,R}(\xi)}\,.
\label{Xdot}
\end{eqnarray}

Simultaneously, at short time scales, the shape fluctuations of the
front around $\phi^{(0)}$ at the instantaneous position of the
Goldstone mode are analyzed by defining the {\it mutually
orthonormal\/} shape fluctuation modes $\{\Psi_m(\xi)\}$ in the
eigenspace of non-zero eigenvalues of the linear stability operator
${\cal L}_{v_{\text{as}}}$ as \cite{review,rocco1}
\begin{eqnarray}
\hspace{0.8cm}\delta\phi(\xi,t)\,=\,\sum_{m\neq0}c_m(t)\,\Psi_{m,R}(\xi)\,.
\label{modes}
\end{eqnarray}
The $c_m(t)$'s are then easily seen to satisfy \cite{review,rocco1}
\begin{eqnarray}
\hspace{-2mm}\dot{c}_m(t)=-\,\tau^{-1}_m\,c_m(t)\,+\,\tilde\varepsilon^{1/2}\int_{-\infty}^{\infty}d\xi\,\Psi_{m,L}(\xi)R(\xi,t),
\label{delphidot}
\end{eqnarray}
where $\Psi_{m,R}(\xi)$ and $\Psi_{m,L}(\xi)$ are respectively the
right and the left eigenvectors of ${\cal L}_{v_{\text{as}}}$ with
eigenvalue $\tau^{-1}_m\neq0$.

If the front position is defined by the position of its Goldstone
mode, then the phenomenon of front diffusion arises due to the random
speed fluctuations $\dot{X}(t)$ of the Goldstone mode around
$v_{\text{as}}$. The diffusion coefficient of the Goldstone mode
(which is identified with the diffusion coefficient of the front
itself in this case) is then easily defined via the Green-Kubo
relation \cite{review,lutz1,rocco1,armero} as
\begin{eqnarray}
D_G\,=\,\frac{\tilde\varepsilon}{2}\,\frac{\int_{-\infty}^{\infty}d\xi\,e^{2v_{\text{as}}\xi}\Phi^2_{G,R}(\xi)\,\langle
g^2[\phi(\xi,t)]\rangle_t}{\left[\int_{-\infty}^{\infty}d\xi\,e^{v_{\text{as}}\xi}\,\Phi^2_{G,R}(\xi)\right]^2}\,.
\label{kuboG}
\end{eqnarray}
The angular brackets with a subscript $t$ denote an average over an
ensemble of (initial) front realizations at time $t$.

On the other hand, if the front position is defined by the position of
its centre of mass (\ref{srt}), then in continuum space, we have
$S(t)=\int_{-\infty}^\infty dx\,\phi(x,t)$, and for reaction-diffusion
fronts (\ref{fkpp}) that satisfy $\phi^{(0)}(\xi)\rightarrow1$ for
$\xi\rightarrow-\infty$, $\phi^{(0)}(\xi)\rightarrow0$ for
$\xi\rightarrow\infty$, and $\delta\phi(\xi)\rightarrow0$ for
$\xi\rightarrow\pm\infty$ \cite{review}
\begin{eqnarray}
\dot{S}(t)=\!\!\underbrace{\int_{-\infty}^\infty
\!\!\!\!d\xi\,\frac{\partial
f(\phi)}{\partial\phi}\bigg|_{\phi^{(0)}(\xi)}\!\!\!\!\delta\phi(\xi,t)}_{b_1(t)}+\underbrace{\tilde\varepsilon^{1/2}\!\!\int_{-\infty}^\infty
\!\!\!\!d\xi\,R(\xi,t)}_{b_2(t)}.
\label{e97b}
\end{eqnarray}
Thereafter [having used Eqs. (\ref{ito2}-\ref{fkpp}), (\ref{modes})
and the solution of (\ref{delphidot})], in the expansion of the
product $\dot{S}(t)\dot{S}(t+t')$ of Eq. (\ref{kubo}),
$b_1(t)b_2(t+t')$ is seen not to contribute to
\begin{eqnarray}
D_f=\frac{1}{2}\lim_{T\rightarrow\infty}\int_{t}^{t+T}\!\!\!\!\!dt'\,\langle\langle\dot{S}(t)\,\dot{S}(t+t')\rangle_\eta\rangle_t,
\label{kubo}
\end{eqnarray}
while $D^{(1)}_f$, $D^{(2)}_f$ and $D^{(3)}_f$, the respective
contributions of $b_2(t)b_2(t+t')$, $b_2(t)b_1(t+t')$ and
$b_1(t)b_1(t+t')$ to $D_f\left(=\sum_{i=1}^3D^{(i)}_f\right)$ are
given by \cite{review}
\begin{widetext}
\begin{eqnarray}
\hspace{3cm}D_f^{(1)}\,=\,D_f^{(2)}\,=\,\frac{\tilde\varepsilon}{2}\int_{-\infty}^\infty
d\xi\,\langle g^2[\phi(\xi,t)]\rangle_t\,,
\label{df2}
\\&&\hspace{-11cm} \mbox{and}\quad
D^{(3)}_f=\sum_{m,m'\neq0}\frac{\tau_m\langle
c_m(t)\,c_{m'}(t)\rangle_t}{2}\underbrace{\int_{-\infty}^\infty
d\xi\,\frac{\delta
f(\phi)}{\delta\phi}\bigg|_{\phi^{(0)}(\xi)}\Psi_{m,R}(\xi)}_{\delta
v_m}\int_{-\infty}^\infty d\xi'\,\frac{\delta
f(\phi)}{\delta\phi}\bigg|_{\phi^{(0)}(\xi')}\Psi_{m',R}(\xi').
\label{df3}
\end{eqnarray}

In general, in this formalism, $g(\phi)$ cannot be replaced by
$g[\phi^{(0)}]$ --- after all, the value of $\phi$ at any given time
depends on its precise evolution history (i.e., noise realization
chosen to evolve the front) at earlier times.  However, in case the
nonlinearities in $f(\phi)$ make $\phi^{(0)}$ pushed, i.e., for
fluctuating pushed fronts with internal noise, the decay times of the
shape fluctuation modes are of the same order as the time scale set by
$1/v_{\text{as}}$. In that case, the dependence of $g(\phi)$ at any
time on the precise evolution history of the front at earlier times
can be neglected to replace $g(\phi)$ by $g[\phi^{(0)}]$. {\it This
replacement makes the stochastic (noise) term in Eq. (\ref{fkpp})
additive in nature. Moreover, it converts Eq. (\ref{delphidot}) to
that of a Brownian particle in a fluid with viscosity $\tau^{-1}_m$,
where the fluctuating force is given by
$\tilde\varepsilon^{1/2}\int_{-\infty}^{\infty}d\xi\,\Psi_{m,L}(\xi)g[\phi^{(0)}(\xi)]\eta$},
implying that one can then use fluctuation-dissipation theorem to
evaluate $\langle
c_m(t)\,c_{m'}(t)\rangle_t=\tilde\varepsilon\int_{-\infty}^{\infty}d\xi\,\Psi_{m,L}(\xi)\Psi_{m',L}(\xi)g^2[\phi^{(0)}(\xi)]/(\tau^{-1}_m+\tau^{-1}_{m'})$
to further obtain \cite{review}
\begin{eqnarray}
D_f^{(3)}=\frac{1}{2}\,D_f^{(1)}\,\,\mbox{i.e.,}\,\,D_f=\frac{5\tilde\varepsilon}{4}\int_{-\infty}^\infty
d\xi\,g^2[\phi^{(0)}(\xi)]\neq
D_G=\frac{\tilde\varepsilon}{2}\,\frac{\int_{-\infty}^{\infty}d\xi\,e^{2v_{\text{as}}\xi}\Phi^2_{G,R}(\xi)\,
g^2[\phi^{(0)}(\xi)]}{\left[\int_{-\infty}^{\infty}d\xi\,e^{v_{\text{as}}\xi}\,\Phi^2_{G,R}(\xi)\right]^2}\,.
\label{dfpushed}
\end{eqnarray}
\end{widetext}

We will see later that for fluctuating ``pulled'' front in the sFKPP
equation too, $D_f$ and $D_G$ are not the same. Before we delve deeper
into the sFKPP equation, here we take a short digression to mention
that similar situation occurs for gas mixtures (see for example,
Chap. 11.2 of Ref. \cite{groot}). Therein, the expression (and the
value) of the diffusion coefficient depends on its precise definition,
but these different expressions of the diffusion coefficient are
(quite non-trivially) related by means of Onsager relations for the
diffusion coefficients. As for fronts too, it should not be surprising
that the precise values of $D_f$ and $D_G$ are not the same. Indeed
what is important to note is that {\it conceptually they are two
entirely different quantities}. Whether they could be related by any
clever means or not is left here for future investigation.

For the fluctuating ``pulled'' front in the sFKPP equation, we make
the ansatz that the front solution can be decomposed to a $\phi^{(0)}$
that is nothing but Brunet and Derrida's cutoff solution, and its
corresponding shape fluctuation modes $\{\Psi_m\}$; i.e., at the
(linearized) leading edge of the front, $\phi^{(0)}(\xi)=\ln
N\sin[q_0(\xi-\xi_1)]e^{-\lambda^*\xi}/\pi$ \cite{bd} and
$\Psi_m(\xi)\simeq\sin[q_m(\xi-\xi_1)]e^{-\lambda^*\xi}/\ln^{1/2}N$
\cite{kns,PvS1}. Here $\lambda^*\!=\!1$, $q_j=(j+1)\pi/\ln
N\,\,\forall j\geq0$, $\xi_1\simeq0$ is the location of the left end
of the leading edge where the nonlinear $\phi^2$ term in the sFKPP
equation is non-negligible, and the cutoff is implemented at
$\xi_0\simeq\ln N$ \cite{bd,PvS1}. This ansatz is consistent with the
compact support property of the front solution in the sFKPP equation
\cite{compact1}, and it also satisfies the requirements for
Eq. (\ref{e97b}) [i.e., we can safely use Eqs. (\ref{df2}-\ref{df3})
with this ansatz].

Then, the $1/\ln^6N$ asymptotic scaling of $D_G$ for the fluctuating
``pulled'' front in the sFKPP equation is obtained very easily
\cite{review}. One simply has to notice that due to the exponential
weight factor in the integrand of the numerator of Eq. (\ref{kuboG}),
the numerator is practically determined within a distance of ${\cal
O}(1/\lambda^*)$ of $\xi_0$, where the integrand scales $\sim\!N$ and
cancels the $1/N$ prefactor. The denominator, on the other hand,
asymptotically simply scales as $\ln^6N$.

As for $D_f$ of the fluctuating ``pulled'' front in the sFKPP
equation, $D^{(1)}_f$ and $D^{(2)}_f$ are straightaway seen to be of
${\cal O}(1/N)$ from Eq. (\ref{df2}), while the evaluation of
$D^{(3)}_f$ is no easy matter. {\it Unlike pushed fronts, one cannot
simply replace $g(\phi)$ by $g[\phi^{(0)}]$ and arrive at
Eq. (\ref{dfpushed}) --- for fluctuating ``pulled'' fronts,
fluctuation modes decay very slowly \cite{kns,PvS1} and as a result,
the front configuration at any time depends strongly on the precise
noise realization that has been used to evolve it at earlier times!\/}
This means that there is no other way forward than to evaluate
$D^{(3)}_f$ in Eq. (\ref{df3}), and this is done in two steps.

At the first step, we argue that in Eq. (\ref{df3}) $\delta
v_m=s_m/\ln^{3/2}N$ with $s_m\sim{\cal O}(1)$ \cite{review} and obtain
\begin{eqnarray}
D^{(3)}_f\,=\,\frac{1}{2}\sum_{m,m'\neq0}\tau_m\frac{s_ms_{m'}}{\ln^3N}\,\langle
c_m(t)\,c_{m'}(t)\rangle_t\,.
\label{e111}
\end{eqnarray}
This scaling of $\delta v_m$ is obtained with the idea that for Brunet
and Derrida's cutoff solution $\phi^{(0)}$, the front speed
$\int_{-\infty}^\infty d\xi\,f[\phi^{(0)}(\xi)]\simeq2$ is of ${\cal
O}(1)$. Naturally, when the front shape $\phi(\xi,t)$ deviates from
$\phi^{(0)}(\xi)$ by an amount $\Psi_{m,R}(\xi)$, which is always a factor
$\ln^{-3/2}N$ weaker than $\phi^{(0)}(\xi)$ itself [notice the
prefactors of $\phi^{(0)}(\xi)$ and $\Psi_{m,R}(\xi)$ above], the
contribution of the $m$-th shape fluctuation mode to the fluctuation
in the front speed $\delta v_m$ has to be of ${\cal O}(\ln^{-3/2}N)$
as well \cite{review}. Furthermore, since there are ${\cal O}(\ln N)$
number of shape fluctuation modes \cite{PvS1,review}, the sum over $m$
and $m'$ in Eq. (\ref{e111}) runs from $1$ to $\ln N$.

At the second (and perhaps the trickiest) step, we determine the
dependence of $\langle c_m(t)\,c_{m'}(t)\rangle_t$ on $\ln N$, and
evaluate the sums in Eq. (\ref{e111}). To this end, notice that for a
given realization, $c_m(t)$ is expressed [via Eq. (\ref{modes})] as
$c_m(t)=\int_{-\infty}^{\infty}d\xi\,\Psi_{m,L}\,\delta\phi(\xi,t)$
\cite{review}. However, the presence of $e^{\lambda^*\xi}$ in
$\Psi_{m,L}(\xi)$ implies that $c_m(t)$ for a given realization is
practically determined from the fluctuation characteristics at the tip
of the front, and therefore, we retain the integral only over the
leading edge of the front:
\begin{eqnarray}
c_m(t)=\frac{1}{\ln^{1/2}N}\int_{\xi_1}^{\xi_0}\!\!d\xi\,e^{\lambda^*\xi}\sin[q_m(\xi-\xi_1)]\,\delta\phi(\xi,t).
\label{e112}
\end{eqnarray}
By virtue of $\langle\delta\phi(\xi,t)\rangle_t=0$, Eq. (\ref{e112})
then yields $\langle c_m(t)\rangle_t=0$ as it should, but in the
absence of any statistics of the shape fluctuations of the front at
time $t$, one cannot obtain an expression of $\langle
c_m(t)\,c_{m'}(t)\rangle_t$ from it. Moreover, {\it since we cannot
replace $g(\phi)$ by $g[\phi^{(0)}]$ for fluctuating ``pulled''
fronts, one cannot use fluctuation-dissipation theorem in
Eq. (\ref{delphidot}) to evaluate $\langle c_m(t)\,c_{m'}(t)\rangle_t$
either --- there is no generic fluctuation-dissipation theorem
available for multiplicative noise \cite{vankampen}}. Nevertheless, we
can still proceed with two approximations. The first one stems from
the fact that although it is clear from Eq. (\ref{e112}) that $c_m(t)$
and $c_{m'}(t)$ ($m\neq m'$) are correlated in general [after all, for
a given realization, all the $c_m(t)$'s are determined through the
{\it same\/} $\delta\phi(t)$], these fluctuation modes will have a
finite correlation ``length'', i.e., $\langle
c_m(t)\,c_{m'}(t)\rangle_t$ will be negligibly small [compared to
$\sqrt{\langle c^2_m(t)\rangle_t\langle c^2_{m'}(t)\rangle_t}$] when
$|m-m'|$ exceeds a certain threshold $a\ll\ln N$. Based on this
anticipation, our approximation is to  choose $a=0$ for the extreme
(and unrealistic) case to simplify the expression for $D^{(3)}_f$ to
(see later for the discussion on non-zero values of $a$)
\begin{eqnarray}
D^{(3)}_f\,=\,\frac{1}{2}\sum_{m\neq0}^{\ln
N}\tau_m\frac{s^2_m}{\ln^3N}\,\langle c^2_m(t)\rangle_t\,.
\label{e113}
\end{eqnarray}
Then the second approximation is that due to the presence of the
$e^{\lambda^*\xi}$ in the integrand of Eq. (\ref{e112}), only the
value of $\delta\phi$ within a distance $\sim1/\lambda^*=1$ of the tip
determines $c_m(t)$. This is seen in the following manner: typically
the magnitude of $\delta\phi(\xi,t)$ is of order
$\sqrt{\phi(\xi,t)/N}$; at the tip, $e^{\lambda^*\xi_0}\sim N$ cancels
$\delta\phi(\xi,t)\sim1/N$ in Eq. (\ref{e112}), but further behind,
the $1/\sqrt{N}$ factor of $\sqrt{\phi(\xi,t)/N}$ can no longer be
compensated by $e^{\lambda^*\xi}$. We therefore use
\begin{eqnarray}
\hspace{-1mm}|c_m(t)|\!\sim\!\!\!\int_{\xi_0-1}^{\xi_0}\!\!\!\!\!d\xi\,\frac{|\sin[q_m(\xi\!-\!\xi_1)]|}{\ln^{1/2}N}\!\sim\!\frac{q_m}{\sqrt{\ln
 N}}\!\sim\!\frac{\pi(m\!+\!1)}{\ln^{3/2}N}.
\label{e114}
\end{eqnarray}
In Eq. (\ref{e114}), $\sin[q_m(\xi\!-\!\xi_1)]$ has been Taylor
expanded around its value zero at $\xi_0\simeq\ln N$. Thereafter, with
$\tau_m=\ln^2N/[\pi^2\{(m+1)^2-1\}]$ \cite{kns,PvS1}, we obtain
\begin{eqnarray}
D_f\simeq D^{(3)}_f\sim\frac{1}{2}\sum_{m\neq0}^{\ln
N}\frac{(m+1)^2\,s^2_m}{[(m+1)^2-1]\ln^4N}\sim\frac{1}{\ln^3N}\,.
\label{e115}
\end{eqnarray}
We end this paper with five final observations: (i) Realistically,
$a\neq0$, but so long as $a\ll\ln N$, which is what one expects in
reality, Eqs. (\ref{df2}), (\ref{e111}), (\ref{e112}) and (\ref{e114})
show that the $1/\ln^3N$ asymptotic scaling of $D_f$ continues to hold.
(ii) We have extensively used the left eigenvector of the stability
operator ${\cal L}_{v_{\text{as}}}$ for reaction-diffusion systems. For
clock model \cite{vanzon}, or for the model that Brunet and Derrida
considered for their simulation \cite{bd3}, construction of the left
eigenvector for the corresponding ${\cal L}_{v_{\text{as}}}$, and
repeating the same exercise (as in here) are nontrivial. Nevertheless,
since all the arguments for the derivation of the $1/\ln^3N$ asymptotic
scaling of $D_f$ in the sFKPP equation are concentrated on the leading
edge (or more precisely, at the very tip of the front) where the
(fluctuating ``pulled'') front properties are model-independent, one
expects to observe the same scaling for $D_f$ for these two models too. In
view of this, the $1/\ln^3N$ asymptotic scaling of $D_f$ seems to be a
generic property of fluctuating ``pulled'' fronts, independent of the
microscopic model. (iii) In the clock model, one can only create non-local
fluctuations in the front shape $\phi$. The ``collisions'' between clocks
are also non-local in nature \cite{vanzon,review}. In reality, however,
these complications matter neither for the front speed, nor for $D_f$ ---
Brunet and Derrida showed \cite{bd3}, in a simplified version of their
original microscopic model (which closely resembles the clock model), that
the localized fluctuations in $\phi$ at the very tip of the front is all
that is needed for the $1/\ln^3N$ asymptotic scaling of $D_f$. (iv) $D_G$
and [through the scaling of the $c_m(t)$'s] $D_f$ are both determined only
from the tip of the front. This is in perfect agreement with Brunet and
Derrida's ``simplified model'' \cite{bd3}. (v) Finally, the scalings of
$D_f$ and $D_G$ have been obtained by means of an ansatz [second paragraph
below Eq. (\ref{dfpushed})]. The integrity of the method used here to
obtain these scalings can be tested by numerically obtaining the scaling
of $D_G$ and the scaling properties of $\langle c_m(t) c_{m'}(t)\rangle_t$
for the FKPP equation. It must also be noted that these numerical
simulations involve extremely high values of $N$, and are notoriously
difficult to perform.

I thank Henk van Beijeren, Jaume Casademunt, W\-o\-u\-t\-e\-r Kager,
Esteban Moro, Bernard Nienhuis, Wim van Saarloos and Ramses van Zon
for useful discussions. D.~P. is financially supported by the Dutch
research organization FOM (Fundamenteel Onderzoek der Materie).


\begin{thebibliography}{03}
\bibitem{strict} Strictly speaking, ${\cal L}_{v_{\text{as}}}$ should
be read as ${\cal L}_{[v_{\text{as}}+\dot{X}(t)]}$, where
$v_{\text{as}}+\dot{X}(t)$ is the instantaneous speed of the Goldstone
mode w.r.t. the laboratory frame at time $t$. However, since
$\dot{X}(t)\ll v_{\text{as}}$, we replace ${\cal
L}_{[v_{\text{as}}+\dot{X}(t)]}$ by ${\cal L}_{v_{\text{as}}}$.
\end{thebibliography}
\end{document}